\documentclass[prd,showpacs,showkeys,amsmath,amssymb,amsfonts,nofootinbib]{revtex4}
\usepackage{latexsym}
\usepackage[dvips]{graphicx}
\usepackage{bm}

\def\ber{\begin{eqnarray}}
\def\eer{\end{eqnarray}}
\def\beq{\begin{equation}}
\def\eeq{\end{equation}}

\begin{document}

\title{Gravitational Lensing and $f(R)$ theories in
the Palatini approach}

\author{Matteo Luca RUGGIERO}
\email{matteo.ruggiero@polito.it} \affiliation{\footnotesize
Dipartimento di Fisica, \\ Politecnico di Torino,  Corso Duca
degli Abruzzi 24, 10129 Torino \\ and INFN, Sezione di Torino}

\date{\today}

\begin{abstract}
We investigate gravitational lensing in the Palatini approach to
the $f(R)$ extended theories of gravity. Starting from an exact
solution of the $f(R)$ field equations, which corresponds to the
Schwarzschild-de Sitter metric and, on the basis of recent studies
on this metric, we focus on some lensing observables, in order to
evaluate the effects of the non linearity of the gravity
Lagrangian. We give estimates for some astrophysical events, and
show that these effects are tiny for galactic lenses, but become
interesting for extragalactic ones.
\end{abstract}

\pacs{95.62.Sb, 04.50.Kd, 95.30.Sf}
\keywords{Gravitational Lensing, Extended Theories of Gravity}

\maketitle

%----------------------------Section-------------------------------
\section{Introduction} \label{sec:intro}
%----------------------------Section-------------------------------

It is well known and accepted nowadays that the Universe is
undergoing a phase of accelerated expansion \cite{uniexp}. General
Relativity (GR), in its standard formulation, cannot explain this
fact without requiring the existence of the so called dark energy,
or introducing a  cosmological constant which, in turn, brings
about other problems, concerning its nature and origin
\cite{peebles03}. According to a different viewpoint, current
observations might suggest the failure of GR to describe
gravitational interactions at cosmological scale: as a natural
consequence, the claim arises for a modification of the theory of
gravity. In recent years, there has been much interest for the so
called $f(R)$ theories of gravity: in these theories, the
gravitational Lagrangian depends on an arbitrary analytic function
$f$ of the scalar curvature $R$ (see \cite{capozfranc07} and
references therein). These theories are also referred to as
\textit{extended theories of gravity}, since they naturally
generalize GR on a geometric ground: in fact,  when $f(R)=R$ the
action reduces to the usual Einstein-Hilbert action, and
Einstein's theory is obtained. The field equations of these
theories can be obtained in the metric formalism \cite{metric}, by
varying the action  with respect to metric tensor, or in the
Palatini formalism, where the action is varied with respect to the
metric and the affine connection, which are supposed to be
independent from each other \cite{palatini}. The increasing
interest for these theories is mainly due to the fact that it
seems that they are able to provide cosmologically viable models
(see \cite{cosmoviable} and references therein), in which both the
late time acceleration and the inflation phase are present.

However, we must remember that even though there are problems in
explaining the cosmological dynamics, GR is in excellent agreement
with other gravity tests, such as Solar System ones and those
coming from binary pulsar observations \cite{will}; hence, as a
matter of fact, every modified theory of gravity should reproduce
GR in a suitable weak field limit. This issue is  crucial for
$f(R)$ theories:  their weak field limits were largely analyzed,
and the comparison with some known tests of gravity was  carried
out. A lively debated developed because some models have been
ruled out \cite{solartestsno}, while other seem to pass Solar
System tests \cite{solartestsok}. Among the other things,  stellar
models have been investigated, and there is not a common agreement
on the results. In fact, on the one hand, the Palatini approach
seems to provide self consistent solutions \cite{stellarok}, on
the other hand, the same problem studied in the metric approach
leads to unacceptable results \cite{stellarno} (however see
\cite{stellarbo }; see \cite{nopal} for general difficulties with
the Palatini approach). Gravitational lensing (i.e. the deflection
of light rays due to the curvature of space-time), besides having
an historical importance, since it gave one of the first
confirmations of Einstein's theory in the famous eclipse of 1919,
is very important today, because data coming from modern lensing
observations are crucial to explore the Universe at very different
scales \cite{petters,schneider,wamb}. It is then natural to ask
whether modifications of the gravity Lagrangian, as those
introduced in the $f(R)$ theories, affect the standard theory of
lensing (where ''standard'' stands for ''GR lensing'') and, in
particular, the interpretation of its phenomenology. Gravitational
lensing in the framework of $f(R)$ was studied in
\cite{capozziello06a}: in particular, theories with $f(R) \propto
R^n$ were investigated in the metric approach, with emphasis on
point-like lenses.
%qui
In the present paper we want to address a similar task, but
working in the Palatini approach (however, as we remark below, our
results apply also to a subset of the solutions metric $f(R)$
gravity). In a previous work \cite{allemandi05} we found an exact
solution of the $f(R)$ field equations for the vacuum case, which
corresponds to the Schwarzschild-de Sitter metric: starting from
this solution and taking into account recent studies on this
metric \cite{rindler07}, we want to investigate the impact of the
non linearity of the gravity Lagrangian on some lensing
observables. The paper is organized as follows: in Section
\ref{sec:frgravity} we briefly review the Palatini approach to
$f(R)$ theories, which lead to the Schwarzschild-de Sitter vacuum
solution; in Section \ref{sec:baslens} we focus on point
like-lensing, in GR (Schwarzschild space-time) and in $f(R)$
theories (Schwarzschild-de Sitter space-time). Finally,
discussions and conclusions are outlined in Section
\ref{sec:numesti}.

%----------------------------Section-------------------------------
\section{A vacuum exact solution of  the $f(R)$ gravity field equations}\label{sec:frgravity}
%----------------------------Section-------------------------------

The equations of motion of  $f(R)$ gravity in the Palatini
formalism can be obtained by independent variations with respect
to the metric and the connection  from the action\footnote{Let the
signature of the $4$-dimensional Lorentzian manifold $M$ be
$(-,+,+,+)$; furthermore,  if not otherwise stated, we use units
such that $G=c=1$.}
\begin{equation}
A=A_{\mathrm{grav}}+A_{\mathrm{mat}}=\int [ \sqrt{g} f (R)+2\kappa
L_{\mathrm{mat}} (\psi, \nabla \psi) ]  \; d^{4}x,
\label{eq:actionf(R)}
\end{equation}
where $R\equiv R( g,\Gamma) =g^{\alpha\beta}R_{\alpha
\beta}(\Gamma )$, $R_{\mu \nu }(\Gamma )$ is the Ricci-like tensor
of any torsionless connection $\Gamma$   independent from the
metric  $g$, which is assumed here to be the physical metric. The
gravitational part of the Lagrangian is represented by any real
analytic function $f (R)$ of  the scalar curvature $R$. The matter
Lagrangian  $L_{\mathrm{mat}}$  is functionally depending on
unspecified matter fields $\Psi$ together with their first
derivatives,
equipped with a gravitational coupling constant $\kappa=\frac{8\pi G}{c^4}$.\\
According to the Palatini formalism \cite{palatini}, from
(\ref{eq:actionf(R)}) we obtain the following equations of motion:
\begin{eqnarray}
f^{\prime }(R) R_{(\mu\nu)}(\Gamma)-\frac{1}{2} f(R)  g_{\mu \nu
}&=&\kappa T_{\mu \nu }^{mat},  \label{ffv1}\\
\nabla _{\alpha }^{\Gamma }[ \sqrt{g} f^\prime (R) g^{\mu \nu })&=&0, \label{ffv2}
\end{eqnarray}
where $T^{\mu\nu}_{mat}=-\frac{2}{\sqrt g}\frac{\delta
L_{\mathrm{mat}}}{\delta g_{\mu\nu}}$ denotes the matter source
stress-energy tensor and $\nabla^{\Gamma}$ means covariant
derivative with respect to the connection $\Gamma$. The equations
of motion (\ref{ffv1}) can be supplemented by the scalar-valued
equation obtained by taking the $g$-trace of (\ref{ffv1}), where
we set $\tau=\mathrm{tr} T=g^{\mu \nu }T^{mat}_{\mu \nu }$:
\begin{equation}
f^{\prime} (R) R-2 f(R)= \kappa \tau.  \label{ss}
\end{equation}
The algebraic equation (\ref{ss}) is called the \textit{structural equation}
and it controls the solutions of equations (\ref{ffv1}).\\

The field equations (\ref{ffv1}-\ref{ffv2}) and the structural
equation (\ref{ss}) in vacuum become
\begin{eqnarray}
[f'(R)] R_{(\mu\nu)}(\Gamma)-\frac{1}{2}[f(R)] g_{\mu \nu
}&=&0 , \label{ffv111}\\
\nabla _{\alpha }^{\Gamma }(\sqrt{g} \; [f'(R)] \; g^{\mu \nu
})&=&0, \label{ffv211} \\
f^{\prime }(R) R-2f(R)&=& 0. \label{eq:fvac2}
\end{eqnarray}
As shown in \cite{allemandi05} (see, in particular, Section 3),
the system of equations (\ref{ffv111}-\ref{eq:fvac2}) has the
spherical symmetrical solution

\begin{equation}
ds^2=-b(r)dt^2+\frac{dr^2}{b(r)} +r^2d\vartheta^2+r^2\sin^2
\vartheta d\varphi^2, \label{eq:metrica1}
\end{equation}
where $b(r) \doteq \left(1-\frac{2M}{r}-\frac{k r^2}{3}\right)$;
$M$ is the mass of the  source of the gravitational field and
$k=-c_i/4$, where $R=c_i$ is any of the solutions of the
structural equation (\ref{eq:fvac2}). In doing so, we have
obtained a solution with constant scalar curvature $R$. In
particular, if $f(R)=R$ (i.e. our theory is GR) then $R=0$ is the
solution of the structural equation, and (\ref{eq:metrica1})
reduces to the classical Schwarzschild solution.\\

\textbf{Remark.\ } We would like to point out that, for a given
$f(R)$ function, in vacuum case the solutions of the field
equations of Palatini $f(R)$ gravity are a subset of the solutions
of the field equations of metric $f(R)$ gravity \cite{metricvspalatini}. In particular
the spherically symmetric solution (\ref{eq:metrica1}) is also a
solution of the field equations of metric $f(R)$ gravity
\textit{with constant scalar curvature $R$}: actually, in metric
$f(R)$ gravity there are also spherically symmetric solutions
where the scalar curvature is not constant (see e.g.
\cite{metricspherically} ). So, strictly speaking, what follows
applies
also to metric $f(R)$ gravity.\\

In GR, the Schwarzschild-de Sitter (\ref{eq:metrica1}) solution
corresponds to a spherically symmetric solution of Einstein field
equations, with a cosmological term $\Lambda g_{\mu\nu}$,
$\Lambda$ being the cosmological constant. In practice, it is
$\Lambda=k$ in our notation. As we said, the cosmological constant
is one of the candidates for explaining the accelerated expansion
of the Universe (in particular, a positive cosmological constant
is required). The data suggest that it cannot exceed the  upper
value of $\Lambda_0 \simeq 10 ^{-52} m^{-2}$ \cite{cosmocost}. As
a consequence, it is reasonable to estimate that the solutions of
the structural equation   have the same order of magnitude; the
same holds for the parameter $k$ in the metric
(\ref{eq:metrica1}).

In a recent paper \cite{rindler07} it has been shown that,
contrary to previous claims (e.g. \cite{lake02}),  the $k$ terms
in (\ref{eq:metrica1}) \textit{does contribute} to the observed
bending of light rays (see also the subsequent papers
\cite{rindler07a}, \cite{sereno07},\cite{lake07},\cite{schuck07}).
The  key point is that the Schwarzschild-de Sitter space-time does
not become flat at spatial infinity and, then, the  spatial
geometry affects deflection measurements.

In our approach, the parameter $k$ is ultimately related to the
non linearity of the action, in fact when $f(R)=R \rightarrow
k=0$. Hence, the study of the effects of the $k$ term on light
bending can be interpreted as an investigation of the effects of
the non linearity of the gravity Lagrangian, which is what we want
to address here.

According to \cite{rindler07}, the bending angle of a light ray in
the gravitational field of a massive object described by
(\ref{eq:metrica1}), turns out to be

\beq \varepsilon_{SdS}=\frac{4M}{\rho} \left( 1 -
\frac{k\rho^4}{24M^2} \right), \label{eq:epssds1}\eeq

up to first order in both $M$ and $k$, if both the source of light
rays and the observer are far from the massive object (the lens).
We see that, when $k=0$, i.e. in GR, $\epsilon_{SdS}$ reduces to
the classical bending angle $\varepsilon_{GR}=\frac{4M}{\rho}$
(e.g. \cite{straumann}). The parameter $\rho$, to lowest order in
$M$ and $k$ is equivalent to the impact parameter. In what
follows, we will investigate gravitational lensing on the basis of
the bending angle (\ref{eq:epssds1}).

%----------------------------Section-------------------------------
\section{Lensing from point-like sources} \label{sec:baslens}
%----------------------------Section-------------------------------

In this Section, we start by briefly reviewing  the basic notions
of gravitational lensing from a spherically symmetrical point-like
source in GR, i.e. we consider lensing in the Schwarzschild field (for a thorough study of Schwarzschild lensing, see e.g. \cite{virb99})
Then, we consider the case of $f(R)$ theories which, on the basis
of what we have said before, corresponds to studying lensing in
the Schwarzschild-de Sitter space-time.

%----------------------------Secchino-------------------------------
\subsection{Basics of lensing from a spherically symmetrical source} \label{ssec:sch}
%----------------------------Section-------------------------------

\begin{figure}[top]
  \begin{center}
    \includegraphics[width=7cm,height=7cm]{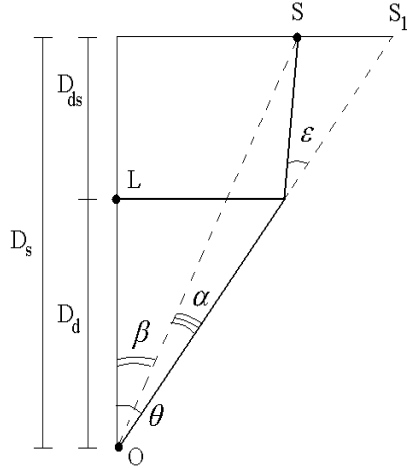}
  \end{center}
  \caption{A simplified lensing geometry: light paths are depicted as straight lines. The observer in $O$ sees the image of the
    source, located at $S$, as if it were in $S_1$. The lens is
    located at the point $L$. Distances between the observer and the source, the observer and the lens and the
    lens and the source are, respectively, defined by $D_s,D_d,D_{ds}$.} \label{fig:lens1}
\end{figure}

In Figure \ref{fig:lens1} a simplified lensing geometry is
depicted. The observer in $O$ sees the image of the source,
located at $S$, as if it were in $S_1$. The lens is located at the
point $L$. In the so called \textit{thin lens approximation} (see
\cite{petters,schneider,wamb}) the light paths are approximated by
straight lines;  if the lens has spherical symmetry, which is what
we assume here, light rays propagate in a plane. We may write the
following relation (see e.g. \cite{virb99}), which links the angular positions of the
images ($\vartheta$), the actual position of the source ($\beta$)
and the bending angle ($\varepsilon$) \beq D_s\tan \vartheta
=D_s\tan \beta +D_{ds}\tan \varepsilon, \label{eq:lens01} \eeq
and, since $D_s \tan \alpha   =D_{ds} \tan \varepsilon $:

\beq D_s\tan \vartheta =D_s\tan \beta +D_{s}\tan \alpha.
\label{eq:lens02} \eeq

So, if we assume that the angles are small (\textit{weak lens
approximation}), we are able to write the following \textit{lens
equation}

\begin{equation}
\beta= \vartheta - \alpha(\vartheta). \label{eq:lens1}
\end{equation}
The latter equation allows to obtain the angular positions
 of the images once those of the sources and the bending angle are known. On using back the fact
that $ \alpha D_s \simeq \varepsilon D_{ds}$, we may write
\begin{equation}
  \beta= \vartheta -\frac{D_{ds}}{D_s} \varepsilon(\vartheta), \label{eq:lens3}
\end{equation}
in which the deflection angle $\varepsilon$ appears.

For instance, in the case of lensing in the Schwarzschild
gravitational field

\begin{equation}
ds^2=-a(r)dt^2+\frac{dr^2}{a(r)} +r^2d\vartheta^2+r^2\sin^2
\vartheta d\varphi^2, \label{eq:metricaS1}
\end{equation}
where $a(r)\doteq\left(1-\frac{2M}{r}\right)$. The point-like lens
is a spherical object of mass $M$; the bending angle turns out to
be $\varepsilon =\varepsilon_{GR}= 4M/\rho$, where $\rho=D_d
\vartheta$ is the impact parameter. On substituting, we get the
explicit form of the lens equation in this case

\begin{equation}
\beta = \vartheta - \frac{D_{ds}}{D_s D_d} \frac{4M}{\vartheta}.
\label{eq:lens4}
\end{equation}

The position corresponding to $\beta=0$, which describes the
alignment of observer, lens and source, defines the so called
\textit{Einstein angle}
\begin{equation}
  \vartheta_E \doteq \sqrt{4M \frac{D_{ds}}{D_s D_d}}, \label{eq:einrad1}
\end{equation}
which allows to define the characteristic length scale of lensing
\beq R_E=\vartheta_E D_d,  \label{eq:einrad2}\eeq which is called
\textit{Einstein radius}. On introducing the Einstein angle in the
lens equation (\ref{eq:lens4}), we get \beq
\beta\vartheta=\vartheta^{2}-\vartheta_E^2, \label{eq:lens41} \eeq or
expressing angles in units of the Einstein angle $\vartheta_E$,
i.e. setting
$\beta=\bar{\beta}\vartheta_E,\vartheta=\bar{\vartheta}\vartheta_E$,
the lens equation has the following expression \beq
\bar{\beta}\bar{\vartheta}=\bar{\vartheta}^2-1. \label{eq:lens42}
\eeq The latter equation can be  simply solved to give the angular
positions of the images \beq \bar{\vartheta}_\pm=
\frac{\bar{\beta}\pm \sqrt{\bar{\beta}^2+4}}{2},
\label{eq:impossds1}\eeq and the angular separation between the
two images \beq \Delta \bar \vartheta \doteq \bar \vartheta_+ -
\bar \vartheta_-=\sqrt{\bar{\beta}^2+4}. \label{eq:impossds2} \eeq

%----------------------------Section-------------------------------
\subsection{Lensing in the Schwarzschild-de Sitter  field} \label{ssec:sds}
%----------------------------Section-------------------------------

In order to obtain the explicit expression of the lens equation in
the Schwarzschild-de Sitter space-time, we have to introduce the
bending angle (\ref{eq:epssds1}) in eq. (\ref{eq:lens3}). So,
taking into account that $\rho=D_d \vartheta$, we may write \beq
\beta=\vartheta-\frac{D_{ds}}{D_sD_d}\frac{4M}{\vartheta}+\frac{D_{ds}D_d^3}{D_s}\frac{\vartheta^3
k}{6M}. \label{eq:lenssds1}   \eeq On introducing the following
dimensionless parameter \beq \chi \doteq
\frac{2}{3}k\frac{D_{ds}^{2}{D_d^{2}}}{D_s^{2}},
\label{eq:defalpha1}\eeq and expressing angles in units of the
Einstein angle $\vartheta_E$ defined by (\ref{eq:einrad1}), the
lens equation becomes \beq
\bar{\beta}\bar{\vartheta}=\bar{\vartheta}^2-1 +\bar{\vartheta}^4
\chi. \label{eq:lenssds2}  \eeq

The lens equation (\ref{eq:lenssds2})  differs from the GR one
(\ref{eq:lens42}) because of the term proportional to $\chi$
which, in turn, is proportional to the parameter $k$,  related to
the non linearity of the gravity Lagrangian. Indeed, we notice
that, beyond the parameter $k$, $\chi$ depends on the geometry of lensing, through the distances
$D_{ds},D_d,D_s$, which enter its definition; $\chi$ is reasonably
small in actual astrophysical events, chiefly because of the
smallness of the parameter $k$, whose expected order of magnitude
is the one of the cosmological constant (see Section
\ref{sec:frgravity}). As a consequence,  the term
$\bar{\vartheta}^4 \chi$ can be treated as a perturbation with
respect to the GR lens equation (\ref{eq:lens42}).

In what follows, we are going to analyze some  observables in
gravitational lensing, in order to evaluate whether the non
linearity of the Lagrangian (\ref{eq:actionf(R)}) raises
detectable effects.

To begin with, we calculate the solutions of the lens equation
(\ref{eq:lenssds2}) for the case $\bar{\beta}=0$, which
corresponds to perfect alignment of observer, lens and source
(superior conjunction): in other words, we look for the modified
Einstein angle  in $f(R)$ theories.

In what follows we  use $\vartheta_{E,GR}$ to refer to the
standard Einstein angle (\ref{eq:einrad1}), i.e. the one obtained
in GR, for a spherically symmetrical point-like lens (described by
the gravitational field (\ref{eq:metricaS1})), while we use
$\vartheta_{E,ETG}$ to indicate the Einstein angle in $f(R)$ for a
spherically symmetrical point-like lens (described by the field
(\ref{eq:metrica1})). Furthermore, we write angles in units of
$\vartheta_{E,GR} \doteq \vartheta_E$, defined in
(\ref{eq:einrad1}).

We look for a solution of the equation \beq 0=\bar{\vartheta}^2-1
+\bar{\vartheta}^4 \chi, \label{eq:lenssdsbetazero1}  \eeq in the
form \beq \bar{\vartheta}_{E,ETG}=1+\delta,
\label{eq:soleqlensbetazero1} \eeq where $\delta $ is assumed to
be small, i.e. of first order in $\chi$. We then obtain \beq
\delta =-\frac{\chi}{2}. \label{eq:soleqlensbetazero2}\eeq Hence
we have, up to first order in $\chi$ \beq
\frac{\vartheta_{E,ETG}-\vartheta_{E,GR}}{\vartheta_{E,GR}}=-\frac{\chi}{2}.
\label{eq:soleqlensbetazero3}\eeq

In particular, we see from (\ref{eq:soleqlensbetazero2}) that the
modified Einstein angle $\vartheta_{E,ETG}$ is smaller (bigger)
than the classical one $\vartheta_{E,GR}$ if $\chi$ is positive
(negative): this means that, according to (\ref{eq:defalpha1}), a
positive (negative) value of $k$ decreases (increases) the
Einstein angle. If we recall that $k$ can be interpreted as a
cosmological constant (see Section \ref{sec:frgravity}), and we
remember that it is required to be positive to mimic the
accelerated expansion of the universe, we conclude that in this
approach to $f(R)$ we obtain an Einstein angle that is smaller
than the standard GR one: this  is indeed independent on the
analytic form of the $f(R)$, and depends on the solutions of the
structural equation only.

We continue our analysis by taking into account the
lensing-induced magnification: actually, gravitational lenses can
amplify the luminosity of the sources, so that faint sources can
become visible. In the case of point-like lenses, the
magnification turns out to be \cite{schneider}

\beq  \textsl A \doteq \left
|\frac{\beta}{\vartheta}\frac{d\beta}{d\vartheta} \right|^{-1}.
\label{eq:defamp1} \eeq

Then, from the lens equation (\ref{eq:lenssds2}) we obtain, up to
first order in $\chi$

\beq \textsl A = \left|1-\frac{1}{\vartheta^4}+2\chi
\left(2\vartheta^{2}-1 \right)\right|^{-1}.
\label{eq:amp1} \eeq

We see that when $\chi=0$ the GR result $A =
\left|1-\frac{1}{\vartheta^4}\right|^{-1}$ is recovered. Actually,
in observations what is seen is the total luminosity, i.e. the sum
of the magnification of each image. However, in practice, it is
not always possible to see the (multiple) images of the sources,
together with their magnification, but the magnification can be
detected thanks to the relative motion of the lens and the source,
that gives rise to a lensing-induced time variability in the
magnification. The time scaling for these variations can be
estimated by $t_E = R_E/v = D_d \vartheta_E/v$, where $v$ is a
typical velocity of the lens (transverse to the line of sight).
Consequently, a variation of the Einstein angle provokes a
variation of this time scale. If we set $t_{ETG}=D_d
\vartheta_{ETG}/v$, according to what we have seen, then, we can
estimate the variation as \beq \frac{\Delta
t}{t_E}=\frac{t_{ETG}-t_{E}}{t_{E}}=\delta=-\frac \chi 2.
\label{eq:deltatesti1} \eeq

Let us now focus on the estimate of the lens mass, determined by
lensing phenomenology: if we suppose that the distances  and the
transverse velocity are known, a measurement of $t_E$ gives an
estimate of the mass $M$ of the lens. As a consequence if we
''incorrectly'' use the Einstein angle of GR instead of the $f(R)$
one, we get a different (biased) mass of the lens. If we denote as
$M$ the ''true'' mas of the lens, and as $M_{GR}$ the one obtained
by using the GR expression of the Einstein angle, we get \beq
\frac{M_{GR}-M}{M}=2\delta =-{\chi}. \label{eq:massestimate1} \eeq
So, according to what we have said above, the effects of $f(R)$ is
an underestimate of the lens mass, in other words the true lens
mass is bigger than the one estimated in standard GR lensing.

Another phenomenological issue that can be investigated is how the
positions of the images (\ref{eq:impossds1}) are modified in
$f(R)$ theories of gravitation. Since $\chi$ is reasonably small,
the lens equation (\ref{eq:lenssds2}) can be solved
perturbatively: in other words we look for the lowest order
corrections of the $k$ term on the GR solutions. To this end, let
$\bar \vartheta_0$ be a solution of the lens equation
corresponding to $\chi=0$: \beq
\bar{\beta}\bar{\vartheta}_0=\bar{\vartheta}_0^2-1.
\label{eq:lenssdsalpha01}  \eeq In particular, $\bar \vartheta_0$
is equal to the solutions (\ref{eq:impossds1}) given above. Then,
we look for solutions in the form \beq \bar \vartheta =\bar
\vartheta_0+\bar \vartheta_1, \label{eq:sollensapprox1} \eeq where
$\bar \vartheta_1$ is supposed to be small. On substituting
(\ref{eq:sollensapprox1}) in (\ref{eq:lenssds2}), and linearizing
in $\bar \vartheta_1$, we obtain, up to first order in $\chi$ \beq
\bar \vartheta_1 = -\frac{\bar \vartheta_0^4 }{2\bar
\vartheta_0-\bar \beta}\chi. \label{eq:sollensapprox2} \eeq  As a
consequence, we may write

\begin{equation}
\bar \vartheta_{\pm} =\bar \vartheta_{0 \pm}-\frac{\bar \vartheta_{0
\pm}^4 }{2\bar \vartheta_{0 \pm}-\bar \beta}\chi.
\label{eq:sollensapprox3}
\end{equation}

If we set $\Delta \bar \vartheta_0=\sqrt{\bar \beta^2+4}$
according to (\ref{eq:impossds2}), we may write the angular
separation between the two images in the form \beq \Delta \bar
\vartheta=\Delta \bar \vartheta_0-g(\Delta \bar \vartheta_0)\chi,
\label{eq:impossds2chi2} \eeq where
\begin{widetext}
\beq g(\Delta \bar \vartheta_0) \doteq \left[{\left(\sqrt{\Delta
\bar \vartheta_0^2-4}+\Delta \bar \vartheta_0
\right)^4}+{\left(\sqrt{\Delta \bar \vartheta_0^2-4}-\Delta \bar
\vartheta_0 \right)^4} \right] \frac{1}{16 {\Delta \bar
\vartheta_0} }. \label{eq:impossds2chi22} \eeq\end{widetext}

In our approach, the modifications (\ref{eq:soleqlensbetazero3}),
(\ref{eq:amp1}), (\ref{eq:deltatesti1}), (\ref{eq:massestimate1}),
(\ref{eq:impossds2chi2}) determined by the non linearity of the
gravity Lagrangian consist in terms that are proportional to the
parameter $\chi$, which is what one would expect in a first order
approximation. As a consequence, the evaluation of the
detectability of these modifications requires an estimate of the
parameter $\chi$  in some actual astrophysical events, which will
be done in the following Section.

\section{Discussion and Conclusions}\label{sec:numesti}

%qui
We have evaluated some effects of the non linearity of the gravity
Lagrangian on lensing phenomenology, in the Palatini approach to
$f(R)$ extended theories of gravity, however our results apply
also to a subset of solutions of metric $f(R)$ gravity.

These effects are controlled by the parameter $\chi$; the latter
is proportional to $k$ which, in turn, is simply related to the
solution of the structural equation (\ref{eq:fvac2}) and it
parameterizes  the deviation of the $f(R)$ theories from GR:
namely, when $k=0$, GR is recovered. The parameter $\chi$ depends
also on the lens geometry, through the distances $D_{ds},D_d,D_s$,
which enter its definition (\ref{eq:defalpha1}); $\chi$ is
expected to be small, because of the smallness of the parameter
$k$, whose  order of magnitude is comparable to the cosmological
constant. Taking into account this fact, we have evaluated the
$f(R)$ effects as perturbations of the standard GR results. To
quantitatively evaluate these effects it is necessary to give
numerical estimates of the dimensionless parameter $\chi$ in some
astrophysical events. Taking into account some reasonable values
for  lenses in our galaxy, we obtain numerically:

\begin{widetext} \beq \chi \simeq 1.5 \times
10^{-10}
\left(\frac{k}{10^{-52}m^{-2}}\right)
\left(\frac{D_{ds}}{10Kpc}\right)^{2}
\left(\frac{D_{d}}{10Kpc}\right)^{2}\left(\frac{20Kpc}{D_{s}}\right)^{2}.
\label{eq:chieval1} \eeq
\end{widetext}

%----------------------------Table-------------------------------
\begin{table}[top]
\begin{center}
\medskip
\begin{tabular}{ccc|c}
$D_d \ [Kpc]$ & $D_s \ [Kpc]$ & $D_{ds} \ [Kpc]$ & $\chi$ \\
\hline
20 & 50  & 30 & $8.6 \times 10^{-10}$\\
 7.6 & 7.6  & $10^{-3}$ & $6.0 \times 10^{-18}$\\
7.6 & 50  & 40 & $2.5 \times 10^{-10}$ \\
 $5 \times 10^3$ & $10^4$ & $10^4$ &$ 1.5 \times 10^{-4}$ \\
\hline
\end{tabular}
\end{center}
\caption{\small Evaluation of the dimensionless parameter $\chi$
(which parameterizes  the deviation of the $f(R)$ theories from
GR) as a function of  the distances between the observer and the
lens, the observer and the source  and the lens and the source,
respectively defined by $D_d,D_s,D_{ds}$ (all measured in $Kpc$);
the value of the parameter $k$ is $10^{-52} \ m^{-2}$.}
\label{tab:table1}
\end{table}
%----------------------------Table-------------------------------

To fix the ideas, we can analyze some specific events. To begin
with, let us consider a lens in the galactic halo, with  $D_d=20 \ Kpc$, and and extra galactic source at $D_s=50
\ Kpc$: we obtain $\chi=8.6 \times 10^{-10}$. Then, we can
consider the supermassive black hole in the radio source Sgr
A$^*$, with  $D_d=7.6 \ Kpc$, and a
source (1) just behind the black hole $D_s=1 \ pc$, (2) outside
the galaxy $D_s=50 \ Kpc$ we get, respectively: (1) $\chi=6.0
\times 10^{-18}$, (2) $\chi=2.5 \times 10^{-10}$.

In general, for sources outside the galaxy we may set $D_{s}\sim
D_{ds}$; consequently if we think of distant galaxies (or cluster
of galaxies) acting as lenses, we obtain numerically

\begin{widetext} \beq \chi \simeq 6.0 \times
10^{-4}
\left(\frac{k}{10^{-52}m^{-2}}\right)
\left(\frac{D_{d}}{10Mpc}\right)^{2}. \label{eq:chieval2} \eeq
\end{widetext}

For instance if we consider $D_d=5 \
Mpc$, we obtain $\chi=1.5 \times 10^{-4}$. These estimates are
summarized in Table \ref{tab:table1}.

% qui
We remember that the order of magnitude of the lensing phenomena
is roughly determined by the Einstein angle; according to eq.
(\ref{eq:soleqlensbetazero3}) modifications of the Einstein angle
due to the non linearity of the gravity Lagrangian are
proportional to $\chi \vartheta_{E,GR}$. As a consequence, the
smallness  of $\chi$  makes the effects of
$f(R)$ theories on lensing hardly detectable for galactic lenses. The situation is different for extragalactic lenses: in fact, in
this case we have $\chi \vartheta_{E,GR} \propto M^{1/2}D_d^{3/2}$,
so that distant galaxies or  clusters of galaxies acting as lenses
can be interesting for our purposes. We notice that similar
results have been obtained in \cite{rindler07a} where, working in
the GR framework, the effects of the cosmological constant on
lensing by distant clusters of galaxies have been evaluated.
However, in the case of extragalactic sources, i.e. when the
lenses are galaxies or clusters of galaxies, the point-like lens
model that we have used in this paper might be oversimplified: 
a more realistic model is needed in order to investigate the possibility of detecting these effects by means of the present
and foreseeable observational techniques.

Nonetheless the estimates that we have obtained suggest at least
the scale where the impact of the $f(R)$-induced corrections on
lensing phenomenology becomes interesting.

To summarize, the non linear terms in the gravity action determine
a modification of  lensing phenomenology; numerically, for
actual astrophysical events, these modifications are very small if
we deal with lenses in our galaxy: these results suggest that the
effect of $f(R)$ are confined around a cosmological scale and,
hence, they are not effective at the
galactic scale. On the contrary, in our simple model, extragalactic
lenses lend themselves as natural candidates for studying the
effects of extended theories of gravity. However, more realistic
models and further studies are needed to evaluate the
detectability of these effects.\\

\textbf{Acknowledgments.} The author acknowledges financial
support from the Italian Ministry of University and Research
(MIUR) under the national program ``Cofin 2005'' - \textit{La
pulsar doppia e oltre: verso una nuova era della ricerca sulle
pulsar}. \\

% --------------------------------------------------------------------------
% The Bibliography
% --------------------------------------------------------------------------

\end{document}